\documentclass[preprint]{elsarticle}
\journal{Parallel Computing}

\usepackage{xcolor}
\usepackage[final]{listings}
\usepackage{amsmath, amssymb}
\usepackage{mathtools}
\usepackage{hyperref}
\usepackage{tikz}
\usepackage{pgfplots}
\usepackage{xspace}
\usepackage{subcaption}
\usepackage{aicescover}

\colorlet{graybg}{gray!10}
\colorlet{plot1}{red!75!white}
\colorlet{plot2}{green!75!black}
\colorlet{plot3}{blue}
\colorlet{plot4}{yellow!75!black}
\colorlet{plot5}{violet}
\colorlet{plot6}{cyan}

\lstdefinelanguage{algs}{
    keywords=[1]{for,each,in,end,set,if,not,function,return},
    keywordstyle=[1]{\bf},
    keywords=[2]{load_start,load_wait,store_start,store_wait,asynchronously,loading,storing},
    keywordstyle=[2]{\bf\green},
    keywords=[3]{copy},
    keywordstyle=[3]{\bf\blue}
}
\usetikzlibrary{
    external,
    scopes,
    patterns,
    calc,
	plotmarks,
    fpu
}
\pgfplotsset{
    every axis/.append style={
        axis background/.style={
            fill=graybg
        },
        legend cell align=left,
        xmin=0,
        ymin=0,
        xlabel near ticks,
        ylabel near ticks,
        enlarge x limits={
            value=0.05,
            auto
        },
        enlarge y limits={
            value=0.05,
            auto
        },
        scaled ticks=false,
        width=\textwidth,
        height=.6\textwidth,
        ymajorgrids=true,
    },
    every axis plot/.append style={
        thick
    },
}
\pgfkeys{
    /tikz/external/mode=list and make
}
\tikzsetexternalprefix{figures/externalized/}
\tikzset{external/export=false}

\let\oldtheequation\theequation
\makeatletter
\def\tagform@#1{\maketag@@@{\ignorespaces#1\unskip\@@italiccorr}}
\renewcommand{\theequation}{(\oldtheequation)}
\makeatother

\hypersetup{
    colorlinks=true,       
    linkcolor=blue!50!black,          
    citecolor=green!50!black,        
    filecolor={.5 0 0},      
    urlcolor=red!50!black,           
}

\newcommand{\red}[1]{\textcolor{red!75!black}{#1}}
\newcommand{\reddesc}{red}
\newcommand{\green}[1]{\textcolor{green!75!black}{#1}}
\newcommand{\greendesc}{green}
\newcommand{\blue}[1]{\textcolor{blue}{#1}}
\newcommand{\bluedesc}{blue}

\newcommand{\splitto}[2]{\biggl(\!\begin{array}{c} #1 \\\hline #2 \end{array}\!\biggr)}
\newcommand{\splitot}[2]{\bigl(#1 \big| #2 \bigr)}
\newcommand{\splittt}[4]{\biggl(\!\begin{array}{c|c} #1 & #2 \\\hline #3 & #4 \end{array}\!\biggr)}

\newcommand{\smpooc}{{\sc SMP-OOC}\xspace}
\newcommand{\smpic}{{\sc SMP-IC}\xspace}
\newcommand{\distooc}{{\sc Elem-1D}\xspace}
\newcommand{\mtdistooc}{{\sc Elem-2D}\xspace}

\aicescovertitle{High Performance Solutions for Big-data GWAS}
\aicescoverauthor{Elmar Peise \and Diego Fabregat-Traver \and Paolo Bientinesi}

\begin{document}
\aicescoverpage

\begin{frontmatter}
    \title{High Performance Solutions for Big-data GWAS}

    \author[aices]{Elmar Peise}
    \ead{peise@aices.rwth-aachen.de}

    \author[aices]{Diego Fabregat-Traver}
    \ead{fabregat@aices.rwth-aachen.de}

    \author[aices]{Paolo Bientinesi}
    \ead{pauldj@aices.rwth-aachen.de}

    \address[aices]{AICES, RWTH Aachen, Schinkelstr. 2, 52062 Aachen, Germany}

    \begin{abstract}
      In order to associate complex traits with genetic polymorphisms,
      genome-wide association studies process huge datasets involving tens of
      thousands of individuals genotyped for millions of polymorphisms.  When
      handling these datasets, which exceed the main memory of contemporary
      computers, one faces two distinct challenges: 1) Millions of
      polymorphisms and thousands of phenotypes come at the cost of hundreds
      of gigabytes of data, which can only be kept in secondary storage; 2)
      the relatedness of the test population is represented by a relationship
      matrix, which, for large populations, can only fit in the combined main
      memory of a distributed architecture.  In this paper, by using
      distributed resources such as Cloud or clusters, we address both
      challenges: The genotype and phenotype data is streamed from secondary
      storage using a double buffering technique, while the relationship
      matrix is kept across the main memory of a distributed memory
      system. With the help of these solutions, we develop separate algorithms
      for studies involving only one or a multitude of traits. We show that
      these algorithms sustain high-performance and allow the analysis of enormous datasets.
    \end{abstract}


\end{frontmatter}


\section{Introduction}
Genome-wide association (GWA) analyses are a powerful statistical tool to
identify certain locations of significance in the genome: Typically, they aim at
determining which {\em single-nucleotide polymorphisms} (SNPs) influences specific
{\em traits} of interest.
Thanks
to these studies, hundreds of SNPs for dozens of complex human diseases and
quantitative traits have been discovered~\cite{pmid19474294}.  
In GWA studies (GWAS),
one of the most used methods to account for the genetic substructure due to
relatedness and population stratification is the variance component approach
based on mixed-models~\cite{pmid3435047,pmid16380716}.  While effective,
mixed-models based methods are computationally demanding both in terms of data
management and computation.  The objective of this research is to make
large-scale GWA analyses affordable.

Computationally, a mixed-model based GWAS on 
$n$ individuals, $m$ genetic markers (SNPs), and $t$ traits 
boils down to the solution of the $m \times t$ generalized least-squares (GLS) problems
\begin{equation}
    \label{eq:GWAS} 
    b_{ij} \coloneqq \bigl( X_i^T M_j^{-1} X_i \bigr)^{-1} X_i^T M_j^{-1} y_j, 
    \ \ \text{with} \ \  i=1, \dots, m \ \ \text{and} \ \ j=1, \ldots, t,
\end{equation}
where 
$X_i \in \mathbb{R}^{n \times p}$ is the design matrix,
$M_j \in \mathbb{R}^{n \times n}$ is the covariance matrix, 
$y_j \in \mathbb{R}^{n}$ contains the vector of observations, and 
$b_{ij} \in \mathbb{R}^{p}$ quantifies
the relation between a variation in an SNP ($X_i$)
and a variation in a trait ($y_j$).  
Furthermore,
$M_j$ is a symmetric positive definite (SPD) matrix,
and the full rank matrix $X_i$
can be viewed as composed of two parts: $X_i = \splitot{X_L}{X_{Ri}}$, with
$X_L \in \mathbb{R}^{n \times (p-1)}$ and $X_{Ri} \in \mathbb{R}^{n \times 1}$,
where $X_L$ is fixed, and only $X_{Ri}$ varies with SNP$_i$.
Moreover, the relationship among the individuals is taken into account by
the covariance matrix $M_j$:
\begin{equation}
    \label{eqn:Mj}
	M_j = \sigma_j^2 \bigl( h_j^2 \Phi + (1 - h_j^2) I \bigr).
\end{equation}
Here, $I$ is the identity matrix, the {\it kinship} matrix $\Phi \in R^{n \times n}$ 
	contains the relationship among all studied individuals, and
$\sigma^2_j$ and $h^2_j$ are trait-dependent scalar estimates.
Finally, common problem sizes are:
$10^3 \le n \le 10^5$, 
$2 \le p \le 20$, 
$10^5 \le m \le 10^8$, and
$t$ is either 1 (single-trait analysis) or in the range of thousands (multi-trait analysis).

The first reported GWA study dates back to 2005: $146$ individuals were
genotyped, and about $103{,}000$ SNPs were analyzed with respect to one trait~\cite{firstgwas}.  Since
then, as the catalog of published GWA analyses shows~\cite{gwascatalog,
gwastrend}, the number of publications has increased steadily, up to
$2{,}404$ in 2011 and $3{,}307$ in 2012.  A similar growth can be observed 
in both the population size and the number of SNPs: Across all the GWAS published
in 2012 the studies comprised on average $15{,}471$ individuals, with a maximum of
$133{,}154$, and on average $1{,}252{,}222$ genetic markers, with a maximum of
$7{,}422{,}970$. 
More recently, advances in technology make it affordable to assess
``omics'' phenotypes in large populations, resulting in the challenge
of analyzing (potentially hundreds) of thousands of traits.
From the perspective of Eqs.~\eqref{eq:GWAS}~and~\eqref{eqn:Mj}, these trends
present concrete challenges, especially in terms of memory requirements.  
As $M_j \in R^{n\times n}$ and the $m$ $X_i$'s and $t$ $y_j$'s compete for the main
memory, two distinct scenarios arise: 
1) if $n$ is small enough for $M_j$ to fit in main memory,
the $X_i$'s and $y_j$'s are to be streamed from disk; 
2) if $M_j$ does not fit in main memory,
both data and computation have to be distributed over multiple compute
nodes. In this paper, we present efficient strategies
for utilizing distributed architectures ---such as clusters, Cloud-based
systems, and supercomputers--- to execute single-trait and multi-trait GWA analyses with 
arbitrarily large population size, number of SNPs, and traits.

\paragraph{Related work}
To perform GWA studies, 
there exist several freely available libraries. Among them, we
highlight {\sc GenABEL}, a widely spread framework for statistical
genomics~\cite{GenABEL}, and {\sc FaST-LMM}, a high-performance software
targeting single-trait analyses~\cite{FastLMM}.
More recently, Fabregat et al.~developed OmicABEL ---a package for the {\sc
  GenABEL} suite--- which implements optimized solutions for shared memory
architectures~\cite{1dooc,2dooc}. However, those algorithms do not support
distributed-memory computations, 
and are only applicable
when the kinship matrix fits in the local memory of a single node.

\paragraph{Organization of the paper}
The rest of this paper is structured as follows.
\autoref{sec:1d} is devoted to single-trait GWAS analyses: 
We commence with the discussion of the core algorithm; 
then, we apply out-of-core techniques to make the algorithm feasible 
for an arbitrary numbers of SNPs;
finally, we cover a distributed-memory extension that allows analyses
of large population sizes.
Similarly, \autoref{sec:2d} addresses multi-trait studies:
We first present a second algorithm specifically tailored for the analysis
of multiple traits;
then, we make use of out-of-core and distributed-memory techniques 
to enable analyses of arbitrary size.
We draw conclusions in~\autoref{sec:conclusion}.


\section{Single-Trait GWAS}
\label{sec:1d}
We consider \autoref{eq:GWAS} restricted to the study of a single trait $y$:
\begin{equation}
    \label{eq:GWAS1d}
    b_i \coloneqq \bigl( X_i^T M^{-1} X_i \bigr)^{-1} X_i^T M^{-1} y, \ \ \text{with} \ \  i=1, \dots, m
\end{equation}

\subsection{The Algorithm}
\label{sec:alg}

The standard route to solving one such GLS is to
reduce it to an ordinary least squares problem (OLS)
$$
    b_i = \bigl(\overline X_i^T \overline X_i\bigr)^{-1} \overline y,
$$
through the operations
\begin{lstlisting}
$L L^T \coloneqq M$ !%
!               !(Cholesky factorization)!
$\overline X_i \coloneqq L^{-1} X_i$ !%
!               !(triangular solve)!
$\overline y \coloneqq L^{-1} y$ !%
!               !(triangular solve)!
\end{lstlisting}

The resulting OLS can then be solved by two alternative approaches,
respectively based on the QR decomposition of  $\overline X_i$, and the
Cholesky decomposition of $\overline X_i^T \overline X_i$.  In general, the
QR-based method is numerically more stable; however, in this specific
application, since $\overline X_i^T \overline X_i \in \mathbb R^{p \times p}$
is very small and $X_i$ is typically well conditioned, both approaches are
equally accurate.  In terms of performance, the solution via Cholesky
decomposition is slightly more efficient:
\begin{lstlisting}[firstnumber=last]
$S_i \coloneqq \overline X_i^T \overline X_i$ !%
!               !(symmetric matrix product)!
$\overline b_i \coloneqq \overline X_i^T \overline y$ !%
!               !(matrix times vector)!
$b_i \coloneqq S_i^{-1} \overline b_i$ !%
!               !(linear system via Cholesky)!
\end{lstlisting}
In this paper, we only consider this approach.

\subsubsection{Multiple SNPs}
\label{sec:algopt}
When the six steps for the solution of one OLS are applied to the specific case
of \autoref{eq:GWAS1d}, by taking advantage of the structure of
$X_i$, it is possible to avoid redundant computation.

Plugging $X_i = \splitot{X_L}{X_{Ri}}$ into $\overline X_i \coloneqq L^{-1}
X_i$ (line~{\tt 2}), we obtain 
$$
    \splitot{\overline X_L}{\overline X_{Ri}} 
    \coloneqq \splitot{L^{-1} X_L}{L^{-1} X_{Ri}},
$$
that is, $\overline X_L \coloneqq L^{-1} X_L$, and $\overline X_{Ri} \coloneqq
L^{-1} X_{Ri}$.  These assignments indicate that the quantity $\overline X_L$
can be computed once and reused across all SNPs.

Similarly, for $S_i \coloneqq \overline X_i^T \overline X_i$ (line~{\tt 4}), we
have\footnote{
    The subscript letters $_L$, $_R$, $_T$, and $_B$, respectively stand for Left, Right,
    Top, and Bottom.
}
$$
    \splittt{S_{TL}}{\ast}{S_{BLi}}{S_{BRi}} 
    \coloneqq 
    \splittt{\overline X_L^T \overline X_L}{\ast}
    {\overline X_{Ri}^T \overline X_L}{\overline X_{Ri}^T \overline X_{Ri}},
$$
from which
\begin{eqnarray}
    S_{TL} &\coloneqq &\overline X_L^T \overline X_L \in \mathbb R^{(p - 1) \times (p - 1)}, \nonumber\\
    S_{BLi} &\coloneqq &\overline X_{Ri}^T \overline X_L \in \mathbb R^{1 \times (p - 1)}, \text{ and } \nonumber \\
    S_{BRi} &\coloneqq &\overline X_{Ri}^T \overline X_{Ri} \in \mathbb R. \nonumber
\end{eqnarray}
This indicates that $S_{TL}$, the top left portion of $S_i$, is independent of $i$
and needs to be computed only once.\footnote{
    Since $S_i$ is symmetric, its top-right and bottom-left quadrants are the
    transpose of each other; we mark the top-right quadrant with a $\ast$ to
    indicate that it is never accessed nor computed.
}  Finally, the same idea also applies to $\overline b_i$ (line~{\tt 5}), yielding
the assignments $\overline b_T \coloneqq \overline X_L^T y$ and $\overline
b_{Bi} \coloneqq \overline X_{Ri}^T y$.

\begin{lstlisting}[
    float=t,
    label=alg:base,
    caption={
        Optimized algorithm for single-trait studies.
    }
]
  $L L^T \coloneqq M$
  $\overline X_L \coloneqq L^{-1} X_L$, $\overline y \coloneqq L^{-1} y$
  $S_{TL} \coloneqq \overline X_L^T \overline X_L$, $\overline b_T \coloneqq \overline X_L^T y$
  for $i$ in $\{1, \ldots, m\}$
      $\overline X_{Ri} \coloneqq L^{-1} X_{Ri}$
      $S_{BLi} \coloneqq \overline X_{Ri}^T \overline X_L$
      $S_{BRi} \coloneqq \overline X_{Ri}^T \overline X_{Ri}$
      $\overline b_{Bi} \coloneqq \overline X_{Ri}^T \overline y$
      set $S_i \coloneqq \splittt{S_{TL}}{\ast}{S_{BLi}}{S_{BRi}}$, $\overline b_i \coloneqq \splitto{\overline b_T}{\overline b_{Bi}}$
      $b_i \coloneqq S_i^{-1} \overline b_i$
  end
\end{lstlisting}


The computation for the whole \autoref{eq:GWAS1d} is given in \autoref{alg:base}.
There, by moving all the operations independent of $i$ outside the loop, 
the overall complexity is lowered from $O(n^3 + m n^2 p)$ down to $O(n^3 + m
n^2)$.\footnote{
    Since in most analyses  $m \gg n$, the complexity reduces by a factor of
    $p$, from $O(m n^2 p)$ down to $O(m n^2)$.
}  This algorithm constitutes the basis for the large-scale versions presented in
the next two sections.


    \subsection{Out-of-core}
    \label{sec:1dooc}
    GWA studies often operate on and generate datasets that exceed the main memory
capacity of current computers.  For instance, a study with $n = 20{,}000$
individuals, $m = 10{,}000{,}000$ SNPs, and $p=4$, requires 1.49~TB to store
the input data ($M$ and $X_i$'s), and generates 305~MB of output.\footnote{
    In practice the size of the output is even larger, because along with 
    each $b_i$, a symmetric $p\times p$ matrix is generated.
}  To make large analyses feasible, regardless of the number of SNPs, Fabregat et
al.~proposed an algorithm that uses asynchronous I/O operations to stream $X_{Ri}$ and
$b_i$ from and to secondary storage~\cite{1dooc}.  This extension of \autoref{alg:base} is
described in the following.

\begin{lstlisting}[
    float=t,
    label=alg:ooc,
    caption={
        Out-of-core algoirithm for single-trait studies.
        The $X_{Ri}$ and $b_i$ are streamed from and to disk in blocks.
        Asynchronous I/O operations are highlighted in \green{\greendesc}.
    }
]
$L L^T \coloneqq M$
$\overline X_L \coloneqq L^{-1} X_L$, !%
!           $\overline y \coloneqq L^{-1} y$
$S_{TL} \coloneqq \overline X_L^T \overline X_L$, !%
!           $\overline b_T \coloneqq \overline X_L^T y$
load_start !first! $X_{blk}$
for each $blk$
    load_wait !current! $X_{blk}$
    if not !last! $blk$: load_start !next! $X_{blk}$
    $\overline X_{blk} \coloneqq L^{-1} X_{blk}$
    for $i$ in  $\{1, \ldots, m_{blk}\}$
        set $\overline X_{Ri} \coloneqq \overline X_{blk}[i]$
        $S_{BLi} \coloneqq \overline X_{Ri}^T \overline X_L$, $S_{BRi} \coloneqq \overline X_{Ri}^T \overline X_{Ri}$
        $\overline b_{Bi} \coloneqq \overline X_{Ri}^T \overline y$
        set $S_i \coloneqq \splittt{S_{TL}}{\ast}{S_{BLi}}{S_{BRi}}$, $\overline b_i \coloneqq \splitto{\overline b_T}{\overline b_{Bi}}$
        $b_i \coloneqq S_i^{-1} \overline b_i$
        set $b_{blk}[i] \coloneqq b_i$
    end
    if not !first! $blk$: store_wait !previous! $b_{blk}$
    store_start !current! $b_{blk}$
end
store_wait !last! $b_{blk}$
\end{lstlisting}


In order to avoid any overhead, the vectors $X_{Ri}$ (and $b_i$) are grouped
into blocks $X_{blk}$ (and $b_{blk}$) of size $m_{blk}$, and read (written)
asynchronously using double buffering.  The idea is to logically split the main
memory in two equal regions: While one region is devoted to the block of data that is
currently processed, the other is used to both store the output from the
previous block and load the input for the next one.  Once the computation on
the current block is completed, the roles of the two regions are swapped.  The
algorithm commences by loading the first block of SNPs $X_{blk}$ from disk into
memory; then, while the GLS's corresponding to this block are solved, the next
block of SNPs is loaded asynchronously in the second memory region.
(Analogously, the previous $b_{blk}$ is stored, while the current one is
computed.)  

When dealing with large analyses, an important optimization comes from,
whenever possible, processing multiple SNPs at once: \autoref{alg:ooc} shows
how combining slow vector operations on $X_{Ri}$ together originates
efficient matrix operations on $X_{blk} \in \mathbb R^{n \times m_{blk}}$ (line~{\tt 8}).

\subsubsection{Shared Memory Implementation}
\label{sec:smp}
The shared memory implementation of \autoref{alg:ooc}, here called~\smpooc, makes use of
parallelism in two different ways~\cite{1dooc}.  The operations in lines~{\tt
1} through~{\tt 8} are dominated by Level 3 Basic Linear Algebra Subroutines
({\sc BLAS}) and take full advantage of a multithreaded implementation of {\sc
BLAS} and the Linear Algebra PACKage ({\sc LAPACK}).  By contrast, for the
operations
within the innermost loop (lines~{\tt 11} through~{\tt 14}), which only involve very
small or thin matrices,  BLAS and especially multithreaded BLAS are
less efficient.  Therefore, they are scheduled in parallel using {\sc OpenMP}
in combination with single-threaded {\sc BLAS} and {\sc LAPACK}.

\paragraph{Performance results}
We compile \smpooc, written in C, with the GNU C~compiler (GCC version~4.4.5) and
link to Intel's Math Kernel Library (MKL version~10.3).  All tests are
executed on a system consisting of two six-core Intel X5675 processors, running
at 3.06~GHz, equipped with 32~GB of RAM, and connected to a 1~TB hard disk.

Preliminary measurements showed that changing $p \in \{1, \ldots, 20\}$
results in performance variation on the order of system fluctuations (below $1
\%$).  Therefore $p = 4$, a value encountered in several GWA
studies, is considered throughout all our experiments.

\begin{figure}[t]
    \centering
    \tikzset{external/export=true}

    \begin{tikzpicture}
        \begin{loglogaxis}[
            ymin={},
            ymax=10800,
            xmin={},
            xlabel={$m$},
            legend pos=north west,
            ytick={1, 10, 60, 600, 3600},
            yticklabels={1 sec, 10 secs, 1 min, 10 mins, 1 hour},
            minor ytick={
                2, 3, 4, 5, 6, 7, 8, 9,
                20, 30, 40, 50,
                120, 180, 240, 300, 360, 420, 480, 540,
                1200, 1800, 2400, 3000,
                7200, 10800
            },
        ]
            \draw[plot1, thick, dotted] (axis cs:200000, 0) -- (axis cs:200000, 400) node[anchor=south, black] {32 GB};
            \addplot[plot1, mark=square*] table[x index=0, y index=1] {figures/data/ooc/hp-gwas_12th.dat};
            \addlegendentry{\smpic}
            \addplot[plot3, mark=*] table[x index=0, y index = 1] {figures/data/ooc/ooc-hp-gwas_12th.dat};
            \addlegendentry{\smpooc}
        \end{loglogaxis}
    \end{tikzpicture}

    \caption[Performance of \smpic and \smpooc ($m$)]{
        Performance of single-trait solvers \smpic and \smpooc as a function of $m$.
        $n = 10{,}000$, $p = 4$, and $m$ ranges from $10^3$ to $10^7$.
        The vertical line indicates the limit for the in-core solver \smpic
        imposed by the RAM size.
    }
    \label{fig:1dooc}
    \tikzset{external/export=false}
\end{figure}
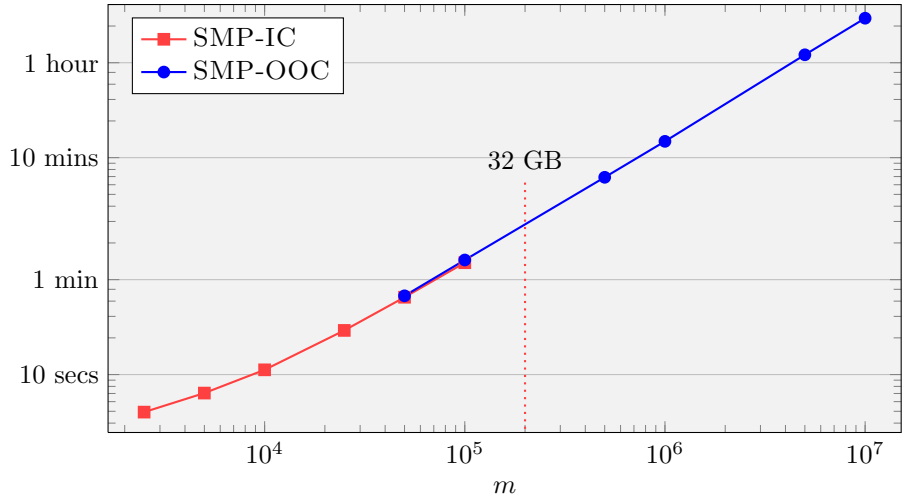


In the first experiment, we compare the efficiency of \smpooc with \smpic, an
equivalent in-core version.  Fixing $n = 10{,}000$, $p = 4$, and we let $m$
vary between $10^3$ and $10^7$.  For the out-of-core version, the SNPs are
grouped in blocks of size $m_{blk} = 5{,}000$.  As \autoref{fig:1dooc} shows,
\smpooc scales linearly in the number of SNPs well beyond the maximum
problem size imposed by the 32~GB of RAM.  Furthermore, the fact that the lines
for the in-core and out-of-core algorithms overlap perfectly confirms that the
I/O operation from and to disk are entirely hidden by computation.

\begin{figure}[t]
    \centering
    \tikzset{external/export=true}

    \begin{tikzpicture}
        \begin{loglogaxis}[
            ymin={},
            xmin={},
            xmax=7e7,
            xlabel={$m$},
            legend pos=south east,
            xtick={1e6,1e7,3.6e7},
            xticklabels={$10^6$,$10^7$,$3.6 \cdot 10^7$},
            ytick={60,3600,86400,604800},
            yticklabels={1 min, 1 hour, 1 day, 1 week},
        ]
            \addplot[plot1, mark=triangle*] table[x index=0, y index=3] {figures/data/ooc/others.dat}
                node[black,anchor=west] {$\times 56.8$};
            \addlegendentry{\sc GenABEL}
            \addplot[plot2, mark=square*] table[x index=0, y index=2] {figures/data/ooc/others.dat}
                node[black,anchor=west] {$\times 6.3$};
            \addlegendentry{\sc FaST-LMM}
            \addplot[plot3, mark=*] table[x index=0, y index=1] {figures/data/ooc/others.dat}
                node[black,anchor=west] {$\times 1$};
            \addlegendentry{\smpooc}
        \end{loglogaxis}
    \end{tikzpicture}

    \caption[Performance of \smpooc, {\sc GenABEL}, and {\sc FaST-LMM}]{
        Performance of the single-trait solver \smpooc compared to {\sc GenABEL} and {\sc FaST-LMM}.
        $n = 10{,}000$, $p = 4$, and $m$ ranges from $10^6$ to $3.6 \cdot 10^7$.
    }
    \label{fig:1dothers}
    \tikzset{external/export=false}
\end{figure}
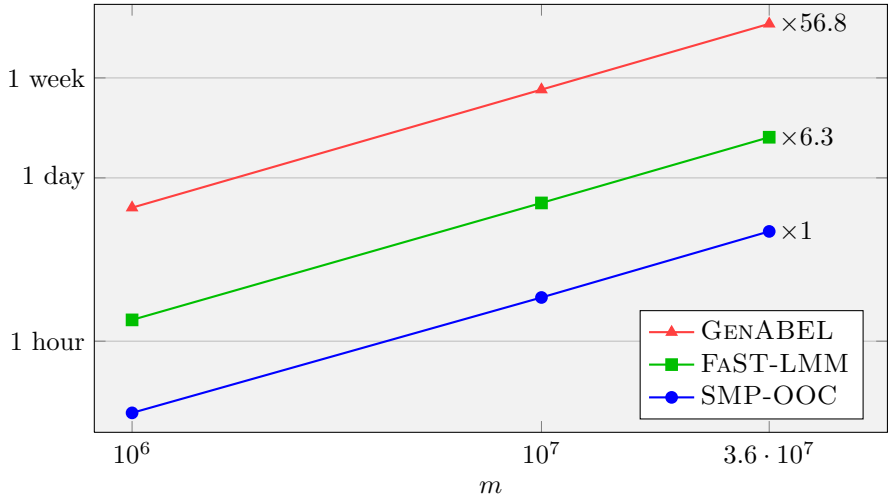


In the second experiment, \autoref{fig:1dothers}, we compare the performance of
\smpooc to that of two other solvers: {\sc FaST-LMM}, a program
designed for GWAS on large datasets~\cite{FastLMM}, and {\sc GenABEL}, a widely
spread library for genome studies~\cite{GenABEL}.  Again, fixing $n =
10{,}000$ and $p = 4$, $m$ varies between $10^6$ and $3.6 \cdot 10^7$.
The fairly constant observed speedups of \smpooc over {\sc FaST-LMM} and {\sc
GenABEL} at $m = 3.6 \cdot 10^7$ are, respectively, $6.3$ and $56.8$.


    \subsection{Distributed Memory}
    \label{sec:1ddist}
    While~\smpooc scales up to an arbitrarily large amount of SNPs $m$, the main
memory is still a limiting factor for the population size $n$: In fact, the
algorithm necessitates the matrix $M \in \mathbb R^{n \times n}$ (or
equivalently, its Cholesky factor $L$) to reside fully in memory.  Due to the
triangular solve (\autoref{alg:ooc}, line~{\tt 2}), keeping the matrix in
secondary storage is not a viable option.  Our approach here consists in
distributing $M$, $L$, and all matrices on which $L$ operates across multiple
compute nodes. Thereby, any constraint on their size is lifted.

\subsubsection{{\sc Elemental}}
\label{sec:elemental}

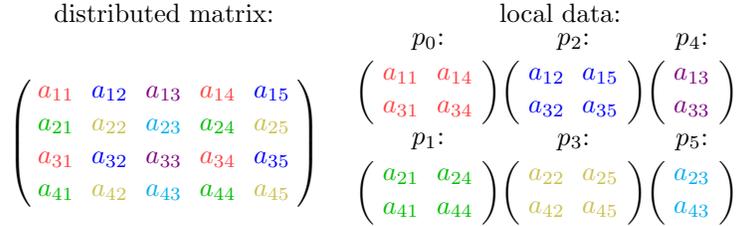
\begin{figure}[t]
    \centering
    \tikzset{external/export=true}

    \begin{tikzpicture}
        \node[matrix, plot1, label={[inner sep=0]above:$p_0$:}] (p0) {
            \node {$a_{11}$}; & \node {$a_{14}$}; \\
            \node {$a_{31}$}; & \node {$a_{34}$}; \\
        };
        \path (p0.south) ++(0, -.2) node[matrix, plot2, anchor=north, label={[inner sep=0]above:$p_1$:}] (p1) {
            \node {$a_{21}$}; & \node {$a_{24}$}; \\
            \node {$a_{41}$}; & \node {$a_{44}$}; \\
        };
        \path (p0.east) ++(.25, 0) node[matrix, plot3, anchor=west, label={[inner sep=0]above:$p_2$:}] (p2) {
            \node {$a_{12}$}; & \node {$a_{15}$}; \\
            \node {$a_{32}$}; & \node {$a_{35}$}; \\
        };
        \path (p1.east) ++(.25, 0) node[matrix, plot4, anchor=west, label={[inner sep=0]above:$p_3$:}] (p3) {
            \node {$a_{22}$}; & \node {$a_{25}$}; \\
            \node {$a_{42}$}; & \node {$a_{45}$}; \\
        };
        \path (p2.east) ++(.25, 0) node[matrix, plot5, anchor=west, label={[inner sep=0]above:$p_4$:}] (p4) {
            \node {$a_{13}$}; \\
            \node {$a_{33}$}; \\
        };
        \path (p3.east) ++(.25, 0) node[matrix, plot6, anchor=west, label={[inner sep=0]above:$p_5$:}] (p5) {
            \node {$a_{23}$}; \\
            \node {$a_{43}$}; \\
        };

        \path (p0.west) -- (p1.west) coordinate[pos=.5] (m);
        \path (m) ++(-.75, 0) node[matrix, anchor=east] (dist) {
            \node[plot1] {$a_{11}$}; & \node[plot3] {$a_{12}$}; & \node[plot5] {$a_{13}$}; & \node[plot1] {$a_{14}$}; & \node[plot3] {$a_{15}$}; \\
            \node[plot2] {$a_{21}$}; & \node[plot4] {$a_{22}$}; & \node[plot6] {$a_{23}$}; & \node[plot2] {$a_{24}$}; & \node[plot4] {$a_{25}$}; \\
            \node[plot1] {$a_{31}$}; & \node[plot3] {$a_{32}$}; & \node[plot5] {$a_{33}$}; & \node[plot1] {$a_{34}$}; & \node[plot3] {$a_{35}$}; \\
            \node[plot2] {$a_{41}$}; & \node[plot4] {$a_{42}$}; & \node[plot6] {$a_{43}$}; & \node[plot2] {$a_{44}$}; & \node[plot4] {$a_{45}$}; \\
        }; 

        \path (p0.north) -- (p4.north) coordinate[pos=.5] (m);
        \path (m) ++(0, .5) node (ld) {local data:};
        \node at (dist |- ld) {distributed matrix:};

        \makeatletter
        \newcommand{\vast}{\bBigg@{4}}
        \makeatother

        \node[inner sep=0] at (dist.west) {$\vast($};
        \node[inner sep=0] at (dist.east) {$\vast)$};

        \foreach \i in {0,...,5} {
            \node[inner sep=0] at (p\i.west) {$\bigg($};
            \node[inner sep=0] at (p\i.east) {$\bigg)$};
        }
    \end{tikzpicture}
    
    \caption{
        Default 2D matrix distribution on a $2 \times 3$ process grid.
    }
    \label{fig:elem2D}
    \tikzset{external/export=false}
\end{figure}


As a framework for distributed-memory dense linear algebra operations, we use
{\sc Elemental}~\cite{elemental}.  This C++ library, which is based on the Message
Passing Interface ({\sc MPI}), operates on a virtual two-dimensional grid
of processes; its name is inspired by the fact that, in general, matrices are
cyclically distributed across this grid in an element-wise
fashion.  This principal distribution\footnote{
    In {\sc Elemental}'s notation: $[MC, MR]$.
}
is shown in~\autoref{fig:elem2D}.

Algebraic operations on distributed matrices typically involve two stages: data
redistribution (communication), and invocation of single-node {\sc BLAS} or
{\sc LAPACK} routines (computation).  Optimal performance is attained by
minimizing communication within the redistributions.  In most cases, as shown
in~\cite{elemental}, this is achieved by choosing the process grid to be as
close to a perfect square as possible.

While a square process grid is optimal for performance, since all processes
only hold non-contiguous portions of the matrix, it complicates loading
contiguously stored data from files into a distributed matrix.  In the context
of GWAS, the algorithm has to load two objects of different nature: the matrix
$M$, and the collections of vectors $X_{blk}$; the special nature of the latter
determines that the vectors can be loaded and processed in any order.

For loading $M$, we first read contiguous panels into the local memory of each
process via standard file operations, and then, by accumulating these panels, construct the global
(distributed) version of $M$.  This is done via {\sc
Elemental}'s axpy-interface, a feature that makes it possible to add node-local
matrices to a global one. 

\begin{figure}[t]
    \centering
    \tikzset{external/export=true}

    \begin{tikzpicture}
        \node[matrix, plot1, label={[inner sep=0]above:$p_0$:}] (p0) {
            \node {$a_{11}$}; & \node {$a_{17}$}; \\
            \node {$a_{21}$}; & \node {$a_{27}$}; \\
            \node {$a_{31}$}; & \node {$a_{37}$}; \\
            \node {$a_{41}$}; & \node {$a_{47}$}; \\
        };
        \path (p0.east) ++(.25, 0) node[matrix, plot3, anchor=west, label={[inner sep=0]above:$p_2$:}] (p1) {
            \node {$a_{12}$}; & \node {$a_{18}$}; \\
            \node {$a_{22}$}; & \node {$a_{28}$}; \\
            \node {$a_{32}$}; & \node {$a_{38}$}; \\
            \node {$a_{42}$}; & \node {$a_{48}$}; \\
        };
        \path (p1.east) ++(.25, 0) node[matrix, plot5, anchor=west, label={[inner sep=0]above:$p_4$:}] (p2) {
            \node {$a_{13}$}; \\
            \node {$a_{23}$}; \\
            \node {$a_{33}$}; \\
            \node {$a_{43}$}; \\
        };
        \path (p2.east) ++(.25, 0) node[matrix, plot2, anchor=west, label={[inner sep=0]above:$p_1$:}] (p3) {
            \node {$a_{14}$}; \\
            \node {$a_{24}$}; \\
            \node {$a_{34}$}; \\
            \node {$a_{44}$}; \\
        };
        \path (p3.east) ++(.25, 0) node[matrix, plot4, anchor=west, label={[inner sep=0]above:$p_3$:}] (p4) {
            \node {$a_{15}$}; \\
            \node {$a_{25}$}; \\
            \node {$a_{35}$}; \\
            \node {$a_{45}$}; \\
        };
        \path (p4.east) ++(.25, 0) node[matrix, plot6, anchor=west, label={[inner sep=0]above:$p_5$:}] (p5) {
            \node {$a_{16}$}; \\
            \node {$a_{26}$}; \\
            \node {$a_{36}$}; \\
            \node {$a_{46}$}; \\
        };

        \path (p0.north) -- (p5.north) coordinate[midway] (m);
        \path (m) ++(0, .66) node[matrix, anchor=south, label={above:distributed matrix:}] (dist) {
            \node[plot1] {$a_{11}$}; & \node[plot3] {$a_{12}$}; & \node[plot5] {$a_{13}$}; & \node[plot2] {$a_{14}$}; & \node[plot4] {$a_{15}$}; & \node[plot6] {$a_{16}$}; & \node[plot1] {$a_{17}$}; & \node[plot3] {$a_{18}$}; \\
            \node[plot1] {$a_{21}$}; & \node[plot3] {$a_{22}$}; & \node[plot5] {$a_{23}$}; & \node[plot2] {$a_{24}$}; & \node[plot4] {$a_{25}$}; & \node[plot6] {$a_{26}$}; & \node[plot1] {$a_{27}$}; & \node[plot3] {$a_{28}$}; \\
            \node[plot1] {$a_{31}$}; & \node[plot3] {$a_{32}$}; & \node[plot5] {$a_{33}$}; & \node[plot2] {$a_{34}$}; & \node[plot4] {$a_{35}$}; & \node[plot6] {$a_{36}$}; & \node[plot1] {$a_{37}$}; & \node[plot3] {$a_{38}$}; \\
            \node[plot1] {$a_{41}$}; & \node[plot3] {$a_{42}$}; & \node[plot5] {$a_{43}$}; & \node[plot2] {$a_{44}$}; & \node[plot4] {$a_{45}$}; & \node[plot6] {$a_{46}$}; & \node[plot1] {$a_{47}$}; & \node[plot3] {$a_{48}$}; \\
        }; 

        \path (m) ++(0, .5) node (ld) {local data:};

        \makeatletter
        \newcommand{\vast}{\bBigg@{4}}
        \makeatother

        \node[inner sep=0] at (dist.west) {$\vast($};
        \node[inner sep=0] at (dist.east) {$\vast)$};

        \foreach \i in {0,...,5} {
            \node[inner sep=0] at (p\i.west) {$\vast($};
            \node[inner sep=0] at (p\i.east) {$\vast)$};
        }
    \end{tikzpicture}
    
    \caption{
        1D matrix distribution on a $1 \times 6$ process grid.
    }
    \label{fig:elem1D}
    \tikzset{external/export=false}
\end{figure}


For loading $X_{blk}$ instead, a collection of contiguously stored vectors is
read into memory through more efficient means than the axpy-interface by
exploiting that, as long as consistently handled, the order of the vectors is
irrelevant.  The trick is to use a matrix that is distributed on a virtual 1D
reordering of the grid into a row of processes.  As shown in
\autoref{fig:elem1D}, the process-local data of such a matrix is a set of full
columns, which can be loaded from a contiguous data-file.  While these local
columns are not adjacent in the distributed matrix, {\sc Elemental} guarantees
that all algebraic operations performed on them maintain their order.  For
performance reasons, prior to any computation, the matrix on the 1D ordering of
this grid needs to be redistributed to conform to the initial 2D process grid
(\autoref{fig:elem2D}).  This redistribution, provided by {\sc Elemental}, can
internally be performed most efficiently through a single {\tt MPI\_Alltoall}
if the 1D grid is the concatenation of the rows of the 2D grid.\footnote{
    In {\sc Elemental}: $[\ast, VR]$.
}

\subsubsection{The Parallel Algorithm}

\begin{lstlisting}[
    float=t,
    label=alg:dist,
    caption={
        Distributed memory algorithm for single-trait studies. 
        Asynchronous I/O operations are depicted \green{\greendesc},
        distributed matrices and operations in \blue{\bluedesc}, and quantities
        that differ across processes in \red{\reddesc}.
    }
]
load_start !first! $\red{X_{blk}}$
$\blue{L L^T \coloneqq M}$
$\blue{\overline X_L \coloneqq L^{-1} X_L}$, !%
!           $\blue{\overline y \coloneqq L^{-1} y}$
copy $\overline X_L\ \blue{\coloneqq \overline X_L}$, $\overline y\ \blue{\coloneqq \overline y}$
$S_{TL} \coloneqq \overline X_L^T \overline X_L$, !%
!           $\overline b_T \coloneqq \overline X_L^T y$
for each $blk$
    load_wait !current! $\red{X_{blk}}$
    if not !last! $blk$: load_start !next! $\red{X_{blk}}$
    set $\blue{X_{blk}} \coloneqq \mathrm{combine}(\red{X_{blk}})$
    $\blue{\overline X_{blk} \coloneqq L^{-1} X_{blk}}$
    set $\red{\overline X_{blk}} \coloneqq \mathrm{localpart}(\blue{\overline X_{blk}})$
    $\red{S_{blk}} \coloneqq \red{\overline X_{blk}}^T \overline X_L$
    for $i$ in $\{1, \ldots, \frac{m_{blk}}{np}\}$
        set $\red{\overline X_{Ri}} \coloneqq \red{\overline X_{blk}}[i]$, $\red{S_{BLi}} \coloneqq \red{S_{blk}}[i]$
        $\red{S_{BRi}} \coloneqq \red{\overline X_{Ri}}^T \red{\overline X_{Ri}}$
        $\red{\overline b_{Bi}} \coloneqq \red{\overline X_{Ri}}^T \overline y$
        set $\red{S_i} \coloneqq \splittt{S_{TL}}{\ast}{\red{S_{BLi}}}{\red{S_{BRi}}}$, $\red{\overline b_i} \coloneqq \splitto{\overline b_T}{\red{\overline b_{Bi}}}$
        $\red{b_i} \coloneqq \red{S_i}^{-1} \red{\overline b_i}$
        set $\red{b_{blk}}[i] \coloneqq \red{b_i}$
    end
    if not !first! $blk$: store_wait !previous! $\red{b_{blk}}$
    store_start !current! $\red{b_{blk}}$
end
store_wait !last! $\red{b_{blk}}$
\end{lstlisting}


In~\autoref{alg:dist}, we present the distributed-memory version
of~\autoref{alg:ooc} for $np$ processes; the matrices distributed
among the processes and the corresponding operations are highlighted
in~\blue{\bluedesc}; the quantities that differ from one process to another are
instead in \red{\reddesc}.

The algorithm begins (line~{\tt 1}) by loading the first $\frac{m_{blk}}{np}$
vectors $X_{Ri}$ into a local block $\red{X_{blk}}$ on each process
asynchronously. Then, from the initially distributed $\blue{M}$,
$\blue{X_L}$, and $\blue{y}$, it computes $\blue{L}$, $\blue{\overline X_L}$,
and $\blue{\overline y}$ (lines~{\tt 2} -- {\tt 3}).  Next, $X_L$ and $y$, respectively
local copies of $\blue{X_L}$ and $\blue{y}$, are created on each
process (line~{\tt 4}).  Since small local computations are significantly more
efficient than the distributed counterparts, all processes compute $S_{TL}$ and $b_T$ 
redundantly (line~{\tt 5}).

In order to compute $\blue{\overline X_{blk} \coloneqq L^{-1} X_{blk}}$,
Elemental requires all involved operands to be distributed across its 2D
process grid.  However, the process-local $\red{X_{blk}}$ is stored as
contiguous columns. These matrices, which are seen as a cyclically distributed
matrix on a 1D~grid (see \autoref{sec:elemental}), are therefore redistributed to
$\blue{X_{blk}}$ on the 2D grid (line~{\tt 9}). After the computation in line~{\tt 10}
completes, the resulting $\blue{X_{blk}}$ is distributed back: Each
process receives those contiguous columns $\red{\overline X_{blk}}$ of
$\blue{\overline X_{blk}}$ that correspond to $\red{X_{blk}}$.

In addition to blocking $\red{X_{Ri}}$ and $\red{b_{Bi}}$, by stacking the
$m_{blk}$ row vectors $\red{S_{BLi}}$'s belonging to the current block into
$\red{S_{blk}}$, their computation is combined into
a single matrix product (line~{\tt 12}).  In line~{\tt 14}, $\red{S_{BLi}}$ and
$\red{X_{Ri}}$ are selected from, respectively, $\red{X_{blk}}$ and
$\red{S_{blk}}$ for the innermost loop.  This loop then computes the local $\red{b_{blk}}$
independently on each process.  Finally, while $\red{b_{blk}}$ (whose columns
$\red{b_i}$ corresponds to the initially loaded vectors $\red{X_{Ri}}$ within
$\red{X_{blk}}$) is stored asynchronously, the next iteration commences.

In addition to {\sc Elemental}'s distributed memory parallelism, we exploit
node-local shared memory parallelism in two different ways: Since The innermost loop
(lines~{\tt 13} through~{\tt 20}) works on very small quantities, it
is parallelized with {\sc OpenMP}; all other operations involve larger
matrices and make use of multithreaded {\sc BLAS} libraries.


        \subsubsection{Performance Results}
        \label{sec:1dperformance}
        We compile \distooc, the C++-implementation of~\autoref{alg:dist}, with the GNU
C++ compiler (GCC version~4.8.1) with {\sc OpenMPI} (version~1.6.4), use {\sc
Elemental} (version~0.82-p1) and link to Intel's Math Kernel Library (MKL
version~11.0).  In our tests, we use a compute cluster with 40 nodes, each
equipped with 16~GB of RAM and two quad-core Intel Harpertown E5450 processors
running at 3.00~Ghz.  The nodes are connected via InfiniBand and access a high
speed Lustre file system.

Throughout all our experiments, we use one process per node with 8 threads
each.  Furthermore, we choose $m_{blk}$ ---the width of $\blue{X_{blk}}$--- as
large as possible to fit in the combined main memory.

\begin{figure}[t]
    \centering
    \tikzset{external/export=true}

    \begin{tikzpicture}
        \begin{loglogaxis}[
            xlabel={$m$},
            ytick={60, 600, 3600, 21600, 86400},
            minor ytick={
                120, 180, 240, 300, 360, 420, 480, 540,
                1200, 1800, 2400, 3000,
                7200, 10800, 14400, 18000,
                43200, 64800,
                172800
            },
            yticklabels={1 min, 10 min, 1 hour, 6 hours, 1 day},
            ymin={},
            xmin={},
            legend pos=north west,
            every axis plot/.append style={
                very thick
            },
        ]
            \draw[plot1, thick, dotted] (axis cs:13858,  1) -- ++(0, 7) node[anchor=south, black] {16 GB};
            \draw[plot2, thick, dotted] (axis cs:37715,  1) -- ++(0, 8) node[anchor=south, black] {32 GB};
            \draw[plot3, thick, dotted] (axis cs:85430,  1) -- ++(0, 9) node[anchor=south, black] {64 GB};
            \draw[plot4, thick, dotted] (axis cs:180860, 1) -- ++(0, 10) node[anchor=south, black] {128 GB};
            \addplot[plot1] file {figures/data/1d/m/1.dat};
            \addlegendentry{8 cores}
            \addplot[plot2] file {figures/data/1d/m/2.dat};
            \addlegendentry{16 cores}
            \addplot[plot3] file {figures/data/1d/m/4.dat};
            \addlegendentry{32 cores}
            \addplot[plot4] file {figures/data/1d/m/8.dat};
            \addlegendentry{64 cores}
        \end{loglogaxis}
    \end{tikzpicture}

    \caption[Performance of \distooc ($m$)]{
        Performance of the single-trait solver \distooc as a function of $m$.
        $n = 30{,}000$, $p = 4$, and $m$ ranges from $10^3$ to $10^7$.
        The vertical lines indicate the limits for in-core versions of the
        parallel algorithm imposed by the combined RAM sizes.
    }
    
    \label{fig:dist:m}
    \tikzset{external/export=false}
\end{figure}
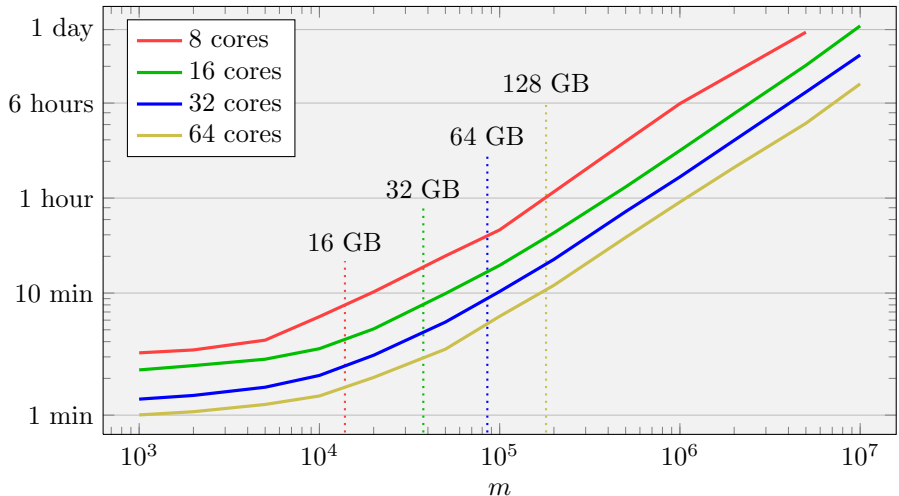

\paragraph{Processing huge numbers of SNPs out-of-core}
Since \distooc incorporates the double-buffering technique
introduced in
\autoref{sec:1dooc}, it can process datasets with arbitrarily large numbers of SNPs $m$ without
introducing any overhead due to I/O operations.
To confirm this claim, we perform a series of experiments on
$np = 1$, $2$, $4$, and $8$ nodes (8, 16, 32, and 64 cores) to solve a problem
of size $n = 30{,}000$ and $p = 4$ with increasing dataset size $m$.  The
performance of these experiments is presented in \autoref{fig:dist:m}, where
the vertical lines mark the points at which the 16~GB of RAM per node are
insufficient to store all $m$ vectors $X_{Ri}$.  The plot shows a very smooth
behavior with $m$ (dominated by the triangular solve in \autoref{alg:dist},
line~{\tt 10}) well beyond this in-core memory limit.

\paragraph{Increasing the population size \texorpdfstring{$n$}{\it n}}
\begin{figure}[t]
    \centering
    \tikzset{external/export=true}

    \begin{tikzpicture}
        \begin{axis}[
            width=.95\textwidth,
            xlabel={$n$},
            ylabel={time [hours]},
            ymax=10,
            xmax=120000,
            legend pos=north west,
            every axis plot/.append style={
                very thick
            },
            xtick={0,20000,40000,60000,80000,100000,120000},
            xticklabels={$0$,$20{,}000$,$40{,}000$,$60{,}000$,$80{,}000$,$100{,}000$,$120{,}000$},
        ]
            \draw[plot1, thick, dotted] (axis cs:46340,  0) -- ++(axis cs:0, 1.1) node[anchor=south, black] {16 GB};
            \draw[plot2, thick, dotted] (axis cs:65536,  0) -- ++(axis cs:0, 2.7) node[anchor=south, black] {32 GB};
            \draw[plot3, thick, dotted] (axis cs:92681,  0) -- ++(axis cs:0, 4.9) node[anchor=south, black] {64 GB};

            \addplot[plot1] file {figures/data/1d/n/1.dat};
            \addlegendentry{8 cores}
            \addplot[plot2] file {figures/data/1d/n/2.dat};
            \addlegendentry{16 cores}
            \addplot[plot3] file {figures/data/1d/n/4.dat};
            \addlegendentry{32 cores}
            \addplot[plot4] file {figures/data/1d/n/8.dat};
            \addlegendentry{64 cores}
        \end{axis}
    \end{tikzpicture}

    \caption[Performance of \distooc ($n$)]{
        Performance of the single-trait solver \distooc as a function of $n$.  
        $p = 4$, $m = 10{,}000$, and $n$ ranges from $5{,}000$ to $120{,}000$.
        The vertical lines indicate the limits imposed by the combined RAM
        sizes.
    }
    \label{fig:dist:n}
    \tikzset{external/export=false}
\end{figure}
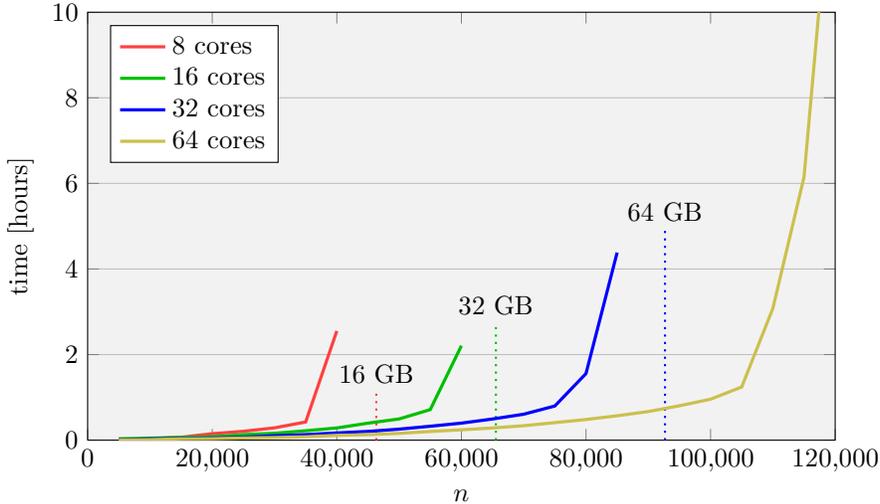


We now turn to the main goal of \distooc: performing computations on problems
whose matrix $M \in \mathbb R^{n \times n}$ exceeds the capacity of the main
memory.  For this purpose, we use $m = 10{,}000$, $p = 4$ and execute \distooc
on $np = 1$, $2$, $4$, and $8$ nodes (8, 16, 32, and 64 cores) with increasing
matrix size $n$.  The performance of these
executions (\autoref{fig:dist:n}) is dominated by the cubic complexity of the Cholesky
factorization of $\blue{M}$ (\autoref{alg:dist}, line~{\tt 2}).  The vertical
lines indicate where the nodes' aggregate main memory would be exceeded by the combined size of the
distributed $\blue{M}$ and the buffers for processing the $X_{Ri}$ one by one.  The plot shows
that our implementation succeeds in overcoming these memory limitations through
increasing the number of nodes.


\section{Multi-Trait GWAS}
\label{sec:2d}
In an important class of GWAS (analysis of ``omics'' phenotypes),
the studies involve many traits $y_j$~\cite{%
gieger08,Hicks2009,Lauc2010,Lauc2013}.
In this case, the set of
generalized least squares problems in \autoref{eq:GWAS} extends into
the second dimension $j$:
\begin{equation}
    \label{eqn:2d}
    b_{ij} \coloneqq \bigl( X_i^T M_j^{-1} X_i \bigr)^{-1} X_i^T M_j^{-1} y_j, \ \ \text{with} \ \  i=1, \dots, m \ \ \text{and} \ \ j=1, \ldots, t.
\end{equation}
This extra dimension is not only reflected in the traits $y_j$, but it also
introduces varying matrices $M_j$.  Such symmetric positive definite $M_j$'s
share the common structure
$$
    M_j = \sigma_j^2 \bigl( h_j^2 \Phi + (1 - h_j^2) I \bigr),
    \label{eqn:2dMj}
$$
where the so called kinship matrix $\Phi \in \mathbb R^{n \times n}$ is
independent of $j$.
This structure plays a critical role in the design of fast algorithms for
multi-trait GWAS.

\subsection{The Algorithm}
\label{2dalg}
In contrast to our single-trait algorithms, which are based on the Cholesky
decomposition $M$, the key to fast algorithms for the multi-trait case is
the eigendecomposition of $\Phi$:
\begin{lstlisting}
$Z \Lambda Z^T \coloneqq \Phi.$
\end{lstlisting}
Here, $Z$ and $\Lambda$ are, respectively, the orthonormal collection of eigenvectors and
the diagonal matrix of eigenvalues of $\Phi$.
Substituting this decomposition, \autoref{eqn:2dMj} becomes
$$
    M_j
    = \sigma_j^2 \bigl( h_J^2 \underbrace{Z^T \Lambda Z}_{\Phi} + (1 - h_j^2) \underbrace{Z^T Z}_{I} \bigr)
    = Z^T \sigma_j^2 \bigl( h_j^2 \Lambda + (1 - h_j^2) I \bigr) Z.
$$
This means that 1) the eigenvectors of $M_j$ and $\Phi$ are the same; 2) the
eigenvalues of $M_j$ are those of $\Phi$ shifted and scaled:
\begin{lstlisting}[firstnumber=last]
$\Lambda_j \coloneqq \sigma_j^2 \bigl( h_j^2 \Lambda + (1 - h_j^2) I \bigr)$
\end{lstlisting}
As a result,\footnote{
    Using the orthonormality of $Z$: $Z^{-1} = Z^T$.
} 
$M_j^{-1}$ can be expressed as
$M_j^{-1} = (Z \Lambda_j Z^T)^{-1} = Z^{-T} \Lambda_j^{-1} Z^{-1} = Z \Lambda^{-1} Z^T$.
Plugging this into \autoref{eqn:2d} yields
\begin{eqnarray}
    b_{ij} &= &\bigl(X_i^T     Z \Lambda_j^{-1}  Z^T X_i  \bigl)^{-1} X_i^T     Z \Lambda_j^{-1}  Z^T y_j  \nonumber\\
           &= &\bigl((Z^T X_i)^T \Lambda_j^{-1} (Z^T X_i) \bigl)^{-1} (Z^T X_i)^T \Lambda_j^{-1} (Z^T y_j).\nonumber
\end{eqnarray}
This expression shows that the assignments
\begin{lstlisting}[firstnumber=last]
$\widetilde X_i \coloneqq Z^T X_i$ !%
!               !(matrix times vector)!
$\widetilde y_j \coloneqq Z^T y_j$ !%
!               !(matrix times vector)!
\end{lstlisting}
can, respectively, be computed independently of $j$ and $i$.
As a result, we have
$$
    b_{ij} = \bigl(\widetilde X_i^T \Lambda_j^{-1} \widetilde X_i)^{-1} \widetilde X_i^T \Lambda_j^{-1} \widetilde y_j.
$$

By decomposing
\begin{lstlisting}[firstnumber=last]
$K_j K_j^T \coloneqq \Lambda_j^{-1}$ !%
!               !(reciprocal square root of diagonal)!
\end{lstlisting}
and assigning
\begin{lstlisting}[firstnumber=last]
$\overline X_{ij} \coloneqq K_j^T \widetilde X_i$ !%
!               !(vector scaling)!
$\overline y_j \coloneqq K_j^T \widetilde y_j,$ !%
!               !(vector scaling)!
\end{lstlisting}
the problem then reduces to 
$$
    b_{ij} = \bigl(\overline X_{ij}^T \overline X_{ij})^{-1} \overline X_{ij}^T \overline y_j.
$$
This ordinary least squares problem is of the same form as encountered
in \autoref{sec:alg}; hence it is solved in the same way.

\begin{lstlisting}[
    float=t,
    label=alg:2dopt,
    caption={
        Optimized algorithm for multi-trait studies.
    }
]
$Z \Lambda Z^T \coloneqq \Phi$
$\widetilde X_L \coloneqq Z^T X_L$
for $i$ in $\{1, \ldots, m\}$
    $\widetilde X_{Ri} \coloneqq Z^T X_{Ri}$
end
for $j$ in $\{1, \ldots, t\}$
    $\widetilde y_j \coloneqq Z^T y_j$
    $\Lambda_j \coloneqq \sigma_j^2 (h_j^2 \Lambda + (1 - h_j^2) I)$
    $K_j K_j^T \coloneqq \Lambda_j^{-1}$
    $\overline X_{Lj} \coloneqq K_j^T \widetilde X_L$, $\overline y_j \coloneqq K_j^T \widetilde y_j$
    $S_{TLj} \coloneqq \overline X_{Lj}^T \overline X_{Lj}$, $\overline b_{Tj} \coloneqq \overline X_L^T \overline y_j$
    for $i$ in $\{1, \ldots, m\}$
        $\overline X_{Rij} \coloneqq K_j^T \widetilde X_{Ri}$
        $S_{BLij} \coloneqq \overline X_{Rij}^T \overline X_{Lj}$, $S_{BRij} \coloneqq \overline X_{Rij}^T \overline X_{Rij}$
        $\overline b_{Bij} \coloneqq \overline X_{Rij}^T \overline y_j$
        set $S_{ij} \coloneqq \splittt{S_{TLj}}{\ast}{S_{BLij}}{S_{BRij}}$, $\overline b_{ij} \coloneqq \splitto{b_{Tj}}{b_{Bij}}$
        $b_{ij} \coloneqq S_{ij}^{-1} \overline b_{ij}$
    end
end
\end{lstlisting}


As in the 1D case, we take advantage of $X_i = \splitot{X_L}{X_{Ri}}$ and
propagate this structure to
$\widetilde X_i = \splitot{\widetilde X_L}{\widetilde X_{Ri}}$,
$\overline X_{ij} = \splitot{\overline X_{Lj}}{\overline X_{Rij}}$,
$\overline S_{ij} = \splittt{\overline S_{TLj}}{\ast}{\overline S_{BLij}}{\overline S_{BRij}}$, and
$\overline b_{ij} = \splitto{\overline b_{Tj}}{\overline b_{Bij}}$.
Extracting all objects independent of the indices $i$ and $j$ from the
corresponding loops, we obtain the mathematically optimized
\autoref{alg:2dopt}.
This optimization reduces the complexity of the algorithm from
$O(n^3 + n^2 (m + t) p + m t n p^2)$ 
to
$O\bigl(n^3 + n^2 (m + t) + m t n p\bigr)$.

\subsection{Out-of-core}
\label{sec:2dooc}

\begin{lstlisting}[
    float=t,
    label=alg:2ddb,
    caption={
        Out-of-core algorithm for multi-trait studies.
        The $y_j$, $X_{Ri}$, and $b_{ij}$ are streamed from and to disk in
        blocks.  Asynchronous I/O operations are highlighted in
        \green{\greendesc}.
        (Function $\mathrm{innerloops}$ is given in \autoref{alg:2dinner}.)
    }
]
$Z \Lambda Z^T \coloneqq \Phi$
$\widetilde X_L \coloneqq Z^T X_L$
for each $blk_m$ !(\green{asynchronously loading} $X_{blk_m}$)!
    $\widetilde X_{blk_m} \coloneqq Z^T X_{blk_m}$
end !(\green{asynchronously storing} $\widetilde X_{blk_m}$)!
for each $blk_t$ !(\green{asynchronously loading} $y_{blk_t}$)!
    $\widetilde y_{blk_y} \coloneqq Z^T y_{blk_t}$
end !(\green{asynchronously storing} $\widetilde y_{blk_t}$)!
for each $tile_t$ !(\green{asynchronously loading} $y_{tile_t}$)!
    for each $tile_m$ !(\green{asynchronously loading} $X_{tile_m}$)!
        $b_{tile} \coloneqq \mathrm{innerloops}(X_{tile_m}, y_{tile_t})$
    end !(\green{asynchronously storing} $b_{tile}$)!
end
\end{lstlisting}


\begin{lstlisting}[
    float=t,
    label=alg:2dinner,
    caption={
        Function $\mathrm{innerloops}$ computes a tile of $b_{ij}$'s from
        corresponding tiles of $X_{Ri}$'s and $y_{j}$'s.
    }
]
function $\mathrm{innerloops}(X_{tile}, y_{tile})$
    for $j$ in $\{1, \ldots, \mathrm{width}(y_{tile})\}$
        set $\widetilde y_j \coloneqq y_{tile}[j]$
        $\Lambda_j \coloneqq \sigma_j^2 (h_j^2 \Lambda + (1 - h_j^2) I)$
        $K_j K_j^T \coloneqq \Lambda_j^{-1}$
        $\overline X_{Lj} \coloneqq K_j^T \widetilde X_L$, $\overline y_j \coloneqq K_j^T \widetilde y_j$
        $S_{TLj} \coloneqq \overline X_{Lj}^T \overline X_{Lj}$, $\overline b_{Tj} \coloneqq \overline X_{Lj}^T \overline y_j$
        for $i$ in $\{1, \ldots, \mathrm{width}(X_{tile})\}$
            set $\widetilde X_{Rj} \coloneqq X_{tile}[i]$
            $\overline X_{Rij} \coloneqq K_j^T \widetilde X_{Ri}$
            $S_{BLij} \coloneqq \overline X_{Rij}^T \overline X_{Lj}$, $S_{BRij} \coloneqq \overline X_{Rij}^T \overline X_{Rij}$
            $\overline b_{Bij} \coloneqq \overline X_{Rij}^T \overline y_j$
            set $S_{ij} \coloneqq \splittt{S_{TLj}}{\ast}{S_{BLij}}{S_{BRij}}$, $\overline b_{ij} \coloneqq \splitto{b_{Tj}}{b_{Bij}}$
            $b_{ij} \coloneqq S_{ij}^{-1} \overline b_{ij}$
            set $b_{tile}[i, j] \coloneqq b_{ij}$
        end
    end
    return $b_{tile}$
end
\end{lstlisting}

To allow processing of arbitrarily large numbers of SNPs $m$ and traits $t$, we
introduce double buffering mechanisms equivalent to those discussed in
\autoref{sec:1dooc}, leading to 
\autoref{alg:2ddb}:
In lines {\tt 3} through {\tt 5}, the matrix-vector products $\widetilde X_{Ri}
\coloneqq Z^T X_{Ri}$ are combined into far more efficient matrix-matrix products on blocks $X_{blk_m}$ of vectors $X_{Ri}$.
While one $\widetilde X_{blk_m}$ is computed, the previous
$\widetilde X_{blk_m}$ and the next $X_{blk_m}$ are, respectively, stored and loaded simultaneously.
Subsequently (lines {\tt 6} -- {\tt 8}), the same mechanism is used to compute
$\widetilde y_j \coloneqq Z^T y_j$ in blocks $y_{blk_t}$.  This process results in two temporary
files containing, respectively, all $\widetilde X_{Ri}$'s and $\widetilde y_j$'s.
These files are of the same size as the inputs $X_{Ri}$ and $y_j$, i.e.,
$n \cdot m$ and $n \cdot t$ doubles.

The main loops of the algorithm (lines {\tt 9} -- {\tt 28}) employ blocking and
double buffering along both $m$ and $t$.  Thereby, the result is computed in
tiles $b_{tile}$ of vectors $b_{ij}$.
While one of these tiles is computed, 
both the next set of vectors $\widetilde X_{Ri}$ (and
$\widetilde y_j$) is loaded in blocks $\widetilde X_{tile_m}$ (and $\widetilde
y_{tile_t}$) and the previous $b_{tile}$ is stored asynchronously.

In total, the algorithm involves four blocking factors corresponding to $blk_t$,
$blk_m$, $tile_t$, and $tile_m$.
In order to make the matrix products involving $y_{blk_t}$ and $X_{blk_m}$ as
efficient as possible, it comes naturally to choose their sizes large.
On the other hand, in order to maximize the computation per IO ratio,
the widths of $y_{tile_t}$ and $X_{tile_m}$ should be chosen such that $b_{tile}$ is roughly
square.

A highly efficient shared memory implementation of \autoref{alg:2ddb} is
presented in~\cite{2dooc}; it is shown to be several orders of
magnitude faster than comparable software packages.

\subsection{Distributed Memory}

Due to the size of the main memory, the kinship matrix $\Phi \in \mathbb R^{n
\times n}$ limits the multi-trait shared memory implementation. (Very much as
the single-trait shared memory implementation was limited by the covariance
matrix $M \in \mathbb R^{n \times n}$.)  To overcome this limitation,  we
present an {\sc Elemental}-based distributed memory solution:
\autoref{alg:2ddist}.

\begin{lstlisting}[
    float=t,
    label=alg:2ddist,
    caption={
        Distributed memory algorithm for multi-trait studies.
        Asynchronous I/O operations are depicted in \green{\greendesc},
        distributed matrices and operations in \blue{\bluedesc}, and quantities
        that differ across processes in \red{\reddesc}.
        (Function $\mathrm{innerloops}$ is given in \autoref{alg:2dinner}.)
    }
]
$\blue{Z} \Lambda \blue{Z^T \coloneqq \Phi}$
$\blue{\widetilde X_L \coloneqq Z^T X_L}$
copy $\widetilde X_L \blue{\coloneqq \widetilde X_L}$
for each $blk_t$ !(\green{asynchronously loading} $\red{y_{blk_t}}$)!
    set $\blue{y_{blk_t}} \coloneqq \mathrm{combine}(\red{y_{blk_t}})$
    $\blue{\widetilde y_{blk_y} \coloneqq Z^T y_{blk_t}}$
    set $\red{\widetilde y_{blk_t}} \coloneqq \mathrm{localpart}(\blue{\widetilde y_{blk_t}})$
end !(\green{asynchronously storing} $\red{\widetilde y_{blk_t}}$)!
for each $blk_m$ !(\green{asynchronously loading} $\red{X_{blk_m}}$)!
    set $\blue{X_{blk_m}} \coloneqq \mathrm{combine}(\red{X_{blk_m}})$
    $\blue{\widetilde X_{blk_m} \coloneqq Z^T X_{blk_m}}$
    set $\red{\widetilde X_{blk_m}} \coloneqq \mathrm{localpart}(\blue{\widetilde X_{blk_m}})$
end !(\green{asynchronously storing} $\red{\widetilde X_{blk_m}}$)!
for each $\red{tile_t}$ !(\green{asynchronously loading} $\red{y_{tile_t}}$)!
    for each $\red{tile_m}$ !(\green{asynchronously loading} $\red{X_{tile_m}}$)!
        $\red{b_{tile}} \coloneqq \mathrm{innerloops}(\red{X_{tile_m}}, \red{y_{tile_t}})$
    end !(\green{asynchronously storing} $\red{b_{tile}}$)!
end
\end{lstlisting}


To overcome the aforementioned limitation,
his algorithm distributes $\Phi$ and its eigenvectors $Z$ across multiple
processes.  Consequently, applying the same
technique used for $X_{blk}$ in the single-trait \autoref{alg:dist}, the blocks
of vectors $X_{blk_m}$ and $y_{blk_t}$ ---to which $\blue{Z}$ is applied---
are also distributed.

The remainder of \autoref{alg:2ddist} (lines~{\tt 14} onward) does not involve
any large matrices that necessitate distributing.  Hence, since each process can work on
separate tiles $\red{b_{tile}}$, this part of the algorithm is embarrassingly
parallel.


    \subsection{Performance Results} 
    The performance experiments for \mtdistooc, the implementation of
\autoref{alg:2ddist} are carried out with the same setup used for \distooc
(\autoref{sec:1dperformance}).

\begin{figure}[t]
    \flushright
    \tikzset{external/export=true}

    \begin{tikzpicture}
        \begin{axis}[
            xlabel={$t$},
            ylabel={time [hours]},
            ymax=5,
            legend pos=north west,
            every axis plot/.append style={
                very thick
            },
        ]

            \addlegendimage{plot1, only marks}
            \addlegendentry{\ 8 cores}
            \addlegendimage{plot2, only marks}
            \addlegendentry{\ 16 cores}
            \addlegendimage{plot3, only marks}
            \addlegendentry{\ 32 cores}
            \addlegendimage{plot4, only marks}
            \addlegendentry{\ 64 cores}

            \addplot[plot1] file {figures/data/2d/tsmall/1.dat};
            \addplot[plot2] file {figures/data/2d/tsmall/2.dat};
            \addplot[plot3] file {figures/data/2d/tsmall/4.dat};
            \addplot[plot4] file {figures/data/2d/tsmall/8.dat};

            \addplot[plot1, dashed, domain=1:100] {384.646 / 60 / 60 * x};
            \addplot[plot2, dashed, domain=1:100] {210.140 / 60 / 60 * x};
            \addplot[plot3, dashed, domain=1:100] {126.855 / 60 / 60 * x};
            \addplot[plot4, dashed, domain=1:100] {86.1381 / 60 / 60 * x};
        \end{axis}
    \end{tikzpicture}

    \tikzset{external/export=false}
    \caption[Performance of \mtdistooc vs \distooc ($t$)]{
        Performance of the multi-trait solver \mtdistooc
        (\tikz[baseline=-.5ex] \draw[thick] (0, 0) -- (.5, 0);)
        compared to $t$ runs of single-trait solver \distooc
        (\tikz[baseline=-.5ex] \draw[thick, dashed] (0, 0) -- (.5, 0);)
        as a function of $t$.
        $n = 30{,}000$, $p = 4$, $m = 10{,}000$, and $t$ ranges from $1$ to
        $100$.  
  }
    
    \label{fig:2dvs1d}
\end{figure}
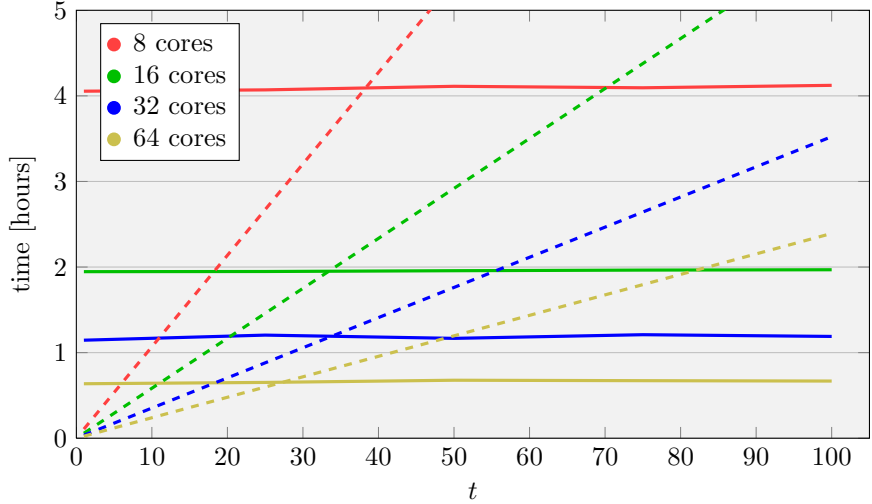


\paragraph{\mtdistooc vs. \distooc}
In the scenario of multi-trait studies ($t>1$), the main advantage of \mtdistooc over
multiple runs of \distooc is that redundant calculations are 
avoided. Complexity-wise, the difference is apparent: $O(m n t p)$ vs.~$O(m n^2
t)$, respectively, for \mtdistooc and \distooc.
The following experiment is designed to illustrate such a gap.
Fixing $n = 30{,}000$, $p = 4$, $m = 10{,}000$, and increasing $t$ from $1$ to $100$, we
compare the runtime for \mtdistooc with that for $t$ independent runs of \distooc.
As \autoref{fig:2dvs1d} shows, regardless of the number
of cores used, \mtdistooc offers the best timings for $t \ge 35$. Most
importantly, the difference in slope indicates that the difference between the
approaches will grow larger as $t$ increases. 
Indeed, at $t = 5000$ \mtdistooc outperforms $t$
executions of \distooc by more than two orders of magnitude.

\begin{figure}[!t]
    \flushright
    \tikzset{external/export=true}

    \begin{tikzpicture}
        \begin{loglogaxis}[
            ymin={},
            xmin={},
            ytick={3600, 21600, 86400},
            minor ytick={
                7200, 10800, 14400, 18000,
                43200, 64800,
                172800
            },
            yticklabels={1 hour, 6 hours, 1 day},
            ymin=1800,
            ymax=259200,
            legend pos=north west,
            every axis plot/.append style={
                very thick
            },
        ]
            \draw[plot1, thick, dotted] (axis cs:6473,   1) -- ++(0, 10) node[anchor=south, black] {16 GB};
            \draw[plot2, thick, dotted] (axis cs:27948,  1) -- ++(0, 10.6) node[anchor=south, black] {32 GB};
            \draw[plot3, thick, dotted] (axis cs:70898,  1) -- ++(0, 11.2) node[anchor=south, black] {64 GB};
            \draw[plot4, thick, dotted] (axis cs:156797, 1) -- ++(0, 12) node[anchor=south, black] {128 GB};

            \addlegendimage{plot1, only marks}
            \addlegendentry{\ 8 cores}
            \addlegendimage{plot2, only marks}
            \addlegendentry{\ 16 cores}
            \addlegendimage{plot3, only marks}
            \addlegendentry{\ 32 cores}
            \addlegendimage{plot4, only marks}
            \addlegendentry{\ 64 cores}

            \addplot[plot1] file {figures/data/2d/m/1.dat};
            \addplot[plot2] file {figures/data/2d/m/2.dat};
            \addplot[plot3] file {figures/data/2d/m/4.dat};
            \addplot[plot4] file {figures/data/2d/m/8.dat};

            \addplot[dashed, plot1] file {figures/data/2d/t/1.dat};
            \addplot[dashed, plot2] file {figures/data/2d/t/2.dat};
            \addplot[dashed, plot3] file {figures/data/2d/t/4.dat};
            \addplot[dashed, plot4] file {figures/data/2d/t/8.dat};
            \coordinate (xlabel) at (xticklabel cs:.5,.5);
        \end{loglogaxis}
        \tikzset{external/export=false}
        \node[anchor=north] at (xlabel) {
            $m$ (\tikz[baseline=-.5ex] \draw[thick] (0, 0) -- (.5cm, 0);)
            /
            $t$ (\tikz[baseline=-.5ex] \draw[thick, dashed] (0, 0) -- (.5cm, 0);)
        };
    \end{tikzpicture}

    \caption[Performance of \mtdistooc ($m$ and  $t$)]{
        Performance of the multi-trait solver \mtdistooc as a function of $m$
        and $t$.
        $n = 30{,}000$, $p = 4$, and, while either $m$ or $t$ is fixed at $10{,}000$,
        the other ranges from $1{,}000$ to $200{,}000$.  
        The vertical lines indicate the limits for in-core versions of the
        parallel algorithm imposed by the combined RAM sizes.
    }
    
    \label{fig:2d.mt}
\end{figure}
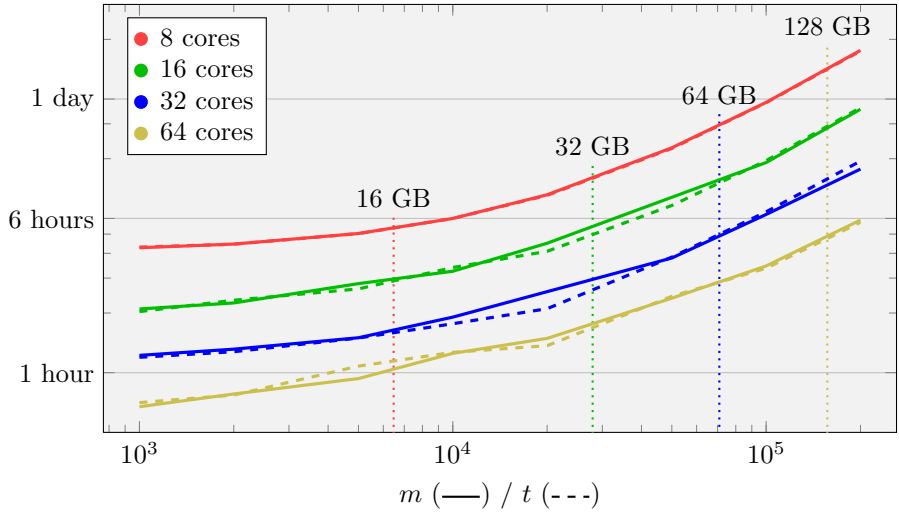


\paragraph{Large \texorpdfstring{$m$ and $t$}{{\it m} and {\it t}}}
Since the aforementioned double-buffering technique is applied to both the inputs ($m$
SNPs $X_{Rj}$ and $t$ traits $y_j$) and the output ($m \cdot t$ vectors
$b_{ij}$), the application of \mtdistooc is not constrained by either the number of SNPs
or the number of traits.  This is shown in \autoref{fig:2d.mt}: In this
experiment, with $n = 30{,}000$ and $p = 4$ constant, one of $m$ and $t$
is fixed to $10{,}000$, and the other varies
between $1{,}000$ and $200{,}000$. The plot provides evidence that 
\mtdistooc solves, without penalty, 
GWAS that do not fit either in ($np = 1$) the local memory of one node (dashed vertical
\textcolor{plot1}{red} line), or ($np > 1$) the
aggregate memory from multiple nodes
(\textcolor{plot2}{green}, \textcolor{plot3}{blue}, and
\textcolor{plot4}{yellow} vertical lines).

In~\autoref{fig:2d.mt},
one should also observe that the timings for the 
experiments with varying $m$, and for those with varying $t$ are very similar.
The reason is that the execution time of \autoref{alg:2ddist}
is dominated by the matrix-matrix multiplications
$\blue{\widetilde X_{blk_m} \coloneqq Z^T X_{blk_m}}$ and $\blue{\widetilde y_{blk_t}
\coloneqq Z^T y_{blk_t}}$ (\autoref{alg:2ddist}, lines~{\tt 6} and~{\tt 11}), 
and the complexity of these operations ---$O(n^2 m + n^2 t)$--- is perfectly symmetric
with respect to $m$ and $t$.

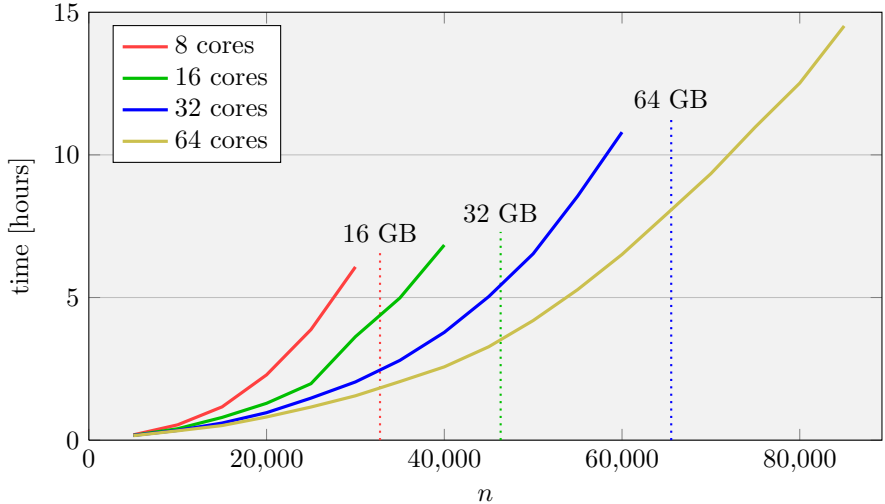
\begin{figure}[t]
    \centering
    \tikzset{external/export=true}

    \begin{tikzpicture}
        \begin{axis}[
            xlabel={$n$},
            ylabel={time [hours]},
            ymax=15,
            legend pos=north west,
            every axis plot/.append style={
                very thick
            },
            xtick={0,20000,40000,60000,80000,100000},
            xticklabels={$0$,$20{,}000$,$40{,}000$,$60{,}000$,$80{,}000$,$100{,}000$},
        ]
            \draw[plot1, thick, dotted] (axis cs:32767, 0) -- ++(axis cs:0, 6.6) node[anchor=south, black] {16 GB};
            \draw[plot2, thick, dotted] (axis cs:46340, 0) -- ++(axis cs:0, 7.3) node[anchor=south, black] {32 GB};
            \draw[plot3, thick, dotted] (axis cs:65535, 0) -- ++(axis cs:0, 11.3) node[anchor=south, black] {64 GB};

            \addplot[plot1] file {figures/data/2d/n/1.dat};
            \addlegendentry{8 cores}
            \addplot[plot2] file {figures/data/2d/n/2.dat};
            \addlegendentry{16 cores}
            \addplot[plot3] file {figures/data/2d/n/4.dat};
            \addlegendentry{32 cores}
            \addplot[plot4] file {figures/data/2d/n/8.dat};
            \addlegendentry{64 cores}
        \end{axis}
    \end{tikzpicture}

    \caption[Performance of \mtdistooc ($n$)]{
        Performance of the multi-trait solver \mtdistooc as a function of $n$.
        $p = 4$, $m = t = 10{,}000$, and $n$ ranges from $5{,}000$ to $85{,}000$.  
        The vertical lines indicate the limits imposed by the combined RAM
        sizes.
    }
    \label{fig:2d.n}
    \tikzset{external/export=false}
\end{figure}


\paragraph{Large \texorpdfstring{$n$}{\it n}}
As the population size $n$ increases, the quadratically growing memory
requirement for the kinship matrix $\Phi \in \mathbb R^{n \times n}$  quickly
surpass the memory available in a single node.
To show that \mtdistooc overcomes this limitation, we fix $m = t =
10{,}000$ and $p = 4$, and let $n$ grow from $5{,}000$ to $85{,}000$. 
The resulting execution times in
\autoref{fig:2d.n} clearly follow a smooth cubic behavior in $n$.\footnote{
    Due to the complexity of the eigenvalue decomposition.
}
As for the single-trait case (\autoref{sec:1ddist}, \autoref{fig:dist:n}),
the plot shows that the limit imposed by the combined main memory size (dashed
vertical lines) can be overcome by increasing the number of nodes.


\section{Conclusion}
\label{sec:conclusion}
We presented parallel algorithms for the computation of
linear mixed-models based 
genome-wide association studies (GWAS).  They address
the issue of growing dataset sizes due to the number of studied polymorphisms
$m$, the population size $n$, and/or the number of traits $t$.

The first algorithm uses a double buffering technique in order to process
datasets with arbitrarily large numbers of genetic polymorphisms.  Compared to
other wide-spread GWAS-codes, our shared memory implementation,
\smpooc, was shown to be at least one order of magnitude faster.

The second algorithm enables the processing of datasets involving large
populations by storing the relationship matrix in the combined main memory of
distributed memory architectures.  \distooc, the implementation of this
algorithm, was shown to scale in both the population size and the
number of processes used.

The third algorithm extends the second by analyzing arbitrary numbers of traits
at once avoiding redundant computation.  This reflects in the performance of
our implementation \mtdistooc, which scales in all problem
sizes, and is significantly faster than multiple runs of \distooc.

Together, these algorithms form a viable basis for the challenges posed by
the scale of current and future genome-wide association studies.

\subsection*{ \bf Acknowledgments}
Financial support from the Deutsche Forschungsgemeinschaft (German Research
Association) through grant GSC 111 is gratefully acknowledged. The authors
thank Yurii Aulchenko for fruitful discussions on the biological background of GWAS.


\bibliographystyle{elsarticle-num}
\bibliography{references}

\end{document}